\documentclass{emulateapj}

\usepackage{amsmath}
\usepackage{amssymb}
\usepackage{graphicx}
\usepackage{color}
\usepackage{multirow}
\usepackage[table]{xcolor}

\newcommand{\Nifs}{\ensuremath{^{56}\mathrm{Ni}}}

\newcommand{\Msun}{\ensuremath{M_{\odot}}}

\newcommand{\rp}{\emph{r}-process}

\received{} 
\accepted{} 
\shortauthors{Barnes and Kasen}
\shorttitle{r-process Light Curves}

\begin{document}

\title{Effect of a High Opacity on the Light Curves of Radioactively Powered Transients from Compact Object Mergers}

\author{Jennifer Barnes$^{1,2}$, Daniel Kasen$^{1,2}$} 
\altaffiltext{1}{Departments of Physics and Astronomy, 366 LeConte Hall, University of California, Berkeley, CA, 94720}
\altaffiltext{2}{Nuclear Science Division, Lawrence Berkeley National Laboratory, Berkeley, CA 94720}

\begin{abstract} 
The coalescence of compact objects are a promising astrophysical
sources of  gravitational wave (GW) signals.  The 
ejection of \rp\ material from such mergers may lead to a
radioactively-powered electromagnetic counterpart which, if
discovered, would enhance the science return of a GW detection.  As very little is known
about the optical properties of heavy \rp\ elements, previous light
curve models have adopted opacities similar to those of iron group
elements.  Here we report that the presence of heavier elements, particularly the
lanthanides, increase the ejecta opacity by several orders of
magnitude.  We include these higher opacities in time dependent,
multi-wavelength radiative transport calculations to predict the
broadband light curves of one-dimensional models over a range of
parameters (ejecta masses $\sim 10^{-3} - 10^{-1} M_{\odot}$ and
velocities $\sim 0.1 -0.3 ~ c$). We find that the higher
opacities lead to much longer duration light curves which can
 last a week or more.  The emission is shifted
toward the infrared bands due to strong optical line
blanketing, and the colors at later times are representative of a
blackbody near the recombination temperature of the lanthanides ($T
\sim 2500$~K).  We further consider the case in which a second mass
outflow, composed of \Nifs, is ejected from a disk wind, and
show that the net result is a distinctive two component spectral
energy distribution, with a bright optical peak due to \Nifs\ and an
infrared peak due to \rp\ ejecta.  We briefly consider the prospects
for detection and identification of these transients.
\end{abstract}


\section{Introduction}

A deeper understanding of compact object binary mergers has the
potential to shed light on several important and unresolved questions
in astrophysics.  Binary neutron star mergers (NSMs) may be the central
engines of short-duration gamma ray bursts \citep{Paczynski_1986_GRB,
  Narayan_Pacz_1992_GRB}, and heavy element nucleosynthesis in merger
ejecta undergoing decompression from nuclear densities
\citep{Lattimer_1974, Lattimer_1976} may contribute to the production
of \rp\ elements in the Universe \citep[e.g.][]{Eichler_1989_rp,
  Rosswog_1998, Freiburghaus_1999, Rosswog_2000, Rosswog_2005,
  Goriely_2005_rp}.  With the LIGO \citep{Abramovici_1992} and VIRGO
experiments preparing to upgrade to ÒadvancedÓ status, compact object mergers
provide the most likely source of kHz gravitational waves (GWs).  The
science returns from detected GW signals can be enhanced by the
identification of a coincident electromagnetic signature
\citep[e.g.][]{Schutz_1986, Schutz_2002, Stubbs_2008, Bloom_2009}.

\citet{Metzger_Berger_2011} consider several possible EM transients
associated with compact object mergers -- short-duration gamma ray bursts, orphan
radio and optical afterglows, and optical ``kilonovae'' light curves
-- and conclude that the last of these offers perhaps the best
opportunity for localizing GW signals.  The kilonovae are optical
transients powered by the radioactive decay of material ejected in the
merger.  Simulations suggest that somewhere in the range of
$10^{-4}-10^{-1}$~\Msun\ may be ejected either by tidal stripping
during the merger itself, or later by a disk wind driven in the evolution
of a post-merger remnant.  The tidally ejected material is very
neutron rich and should form heavy elements via the \rp, while the
subsequent outflows are probably less neutron rich and nucleosynthesis
may not extend much beyond the iron peak.  In either case, the
brightness of the resulting light curves will be set by the mass of
radioisotopes ejected, while the duration depends on the effective
diffusion time through the ejecta, and hence on the mass, velocities,
and opacities.  While the nucleosynthesis and radioactive heating has
been modeled with a fair degree of detail \citep{Metzger_2010,
  Roberts_2011, Goriely_2011}, the unknown opacities of \rp\ elements
have posed a major challenge to predicting kilonova light curves and
spectra.

\citet{Li_Paz_1998} were the first to study the \rp\ transients from
NSM. Their simplified one-zone model assumed spherical symmetry and a
blackbody spectrum, and absorbed much of the physical complexity of
the system into a set of input parameters. Assuming opacities on the
order of the electron scattering opacity, they predicted that the
light curves would peak in the optical/ultraviolet and rise to peak
bolometric luminosities of $10^{42} - 10^{44} ~\rm{ergs ~ s^{-1}}$ on
a timescale of a day.  More recent, more detailed, models
\citep{Metzger_2010, Roberts_2011} have used nuclear reaction networks
to determine the radioactive heating rate and Monte Carlo radiation
transport to calculate merger light curves, under the assumption that
the opacities of heavy \rp\ elements could be approximated by those of
iron. These newer models found qualitatively similar results, with
peak bolometric luminosities $\sim 10^{41} - 10^{42.5} ~ \rm{ ergs ~
  s^{-1}}$ and rise times around one day, with the colors rapidly
reddening post-peak.

In this paper, we show that using more realistic opacities of
\rp\ material has a dramatic effect on the predicted kilonova light
curves.  We use improved estimates of the atomic data of heavy
elements derived from \emph{ab initio} atomic structure models
\citep[][hereafter K13]{Kasen_2013_AS}.  The \rp\ opacities we find are orders of
magnitudes higher than the those of iron group elements; as a
consequence, we predict light curves that are longer, dimmer, and
redder than previously thought. Rather than peaking sharply at $t
\simeq 1~ \mathrm{day}$, the duration of the bolometric light curves
can last $\sim 1$~week.  The spectral energy distribution (SED) is
highly suppressed in the optical, with the bulk of the energy
emitted in the infrared.  Such findings can inform
observational searches for an EM counterpart to a GW trigger by
clarifying the transient timescales, the bands in which EM emission
will be strongest (or have the most distinct signature), and the
distances out to which we might expect a successful EM detection.

\section{Opacity of r-process ejecta}
\label{sec:rpop}

 \begin{figure}
\includegraphics{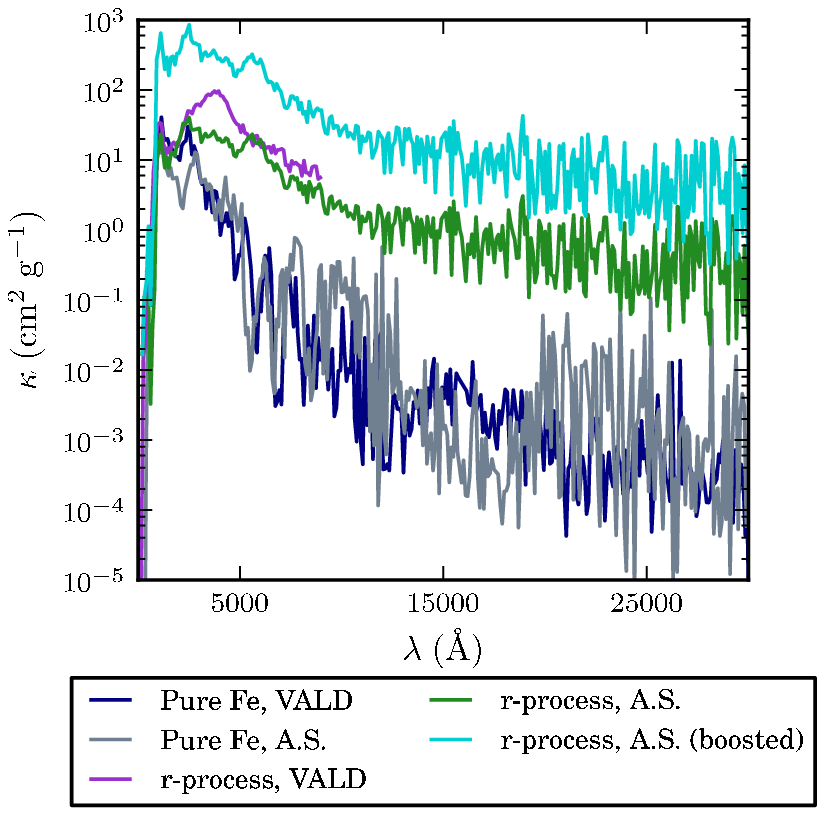}
\vspace{-25pt}
 \caption{Wavelength dependent expansion opacities for ejecta with $\rho = 10^{-13} {\rm ~g ~ cm^{-3},} ~T = 5000 ~{\rm K, ~and ~} t_{\rm exp} = 1 \rm{~day.}$  The opacity of iron is calculated using both the VALD and \texttt{Autostructure} linelists to demonstrate the reliability of the latter approach. The  \rp\ opacity calculated using \texttt{Autostructure} data for Nd is in fairly good agreement with that using the VALD linelist (which only includes extensive line data for a few heavy elements). The \textit{boosted} \rp\ opacity takes into account the diversity of species in an \rp\ mixture by assuming that all lanthanides have an opacity comparable to Nd.}  
 \label{Fig:1}
 \end{figure}
 
Supernova calculations suggest that for complex ions (e.g., the iron
group) bound-bound transitions dominate other forms of opacity, such
as electron scattering, free-free, and photoionization
\citep[e.g.][]{Pinto_2000a}.  Literally millions of lines,
Doppler-broadened by the remnant's differential velocities, will
contribute to a psuedo-continuum bound-bound opacity.  Photons
traveling through the ejecta are continually Doppler-shifted with
respect to the comoving frame, and come into resonance with multiple
transitions one by one. The velocity gradient of the remnant thus
enhances the effective line opacity \citep{Karp_1977}.  We account for
this effect using the expansion opacity formalism introduced by
\citet{Karp_1977} and further developed by \citet{Eastman_1993}, where
the extinction coefficient is given by
\begin{equation}
\alpha_{\rm exp}(\lambda_c) = \frac{1}{ct_{\rm exp}} \sum\limits_i \frac{\lambda_i}{\Delta\lambda_c}(1 - e^{-\tau_i}).
\label{eq:alpha}
\end{equation}
This formula represents an average over discrete wavelength bins,
where $t_{\rm exp}$ is the time since mass ejection, $\lambda_c$ is the central wavelength of the bin,
$\Delta\lambda_c$ is the bin width, $\tau_i$ is the Sobolev optical
depth of a line \citep{Sobolev_1960}, and the sum runs over all lines
in the bin. The extinction coefficient is related to the expansion
opacity by $\kappa_{\rm exp} = \alpha_{\rm exp}/\rho$, where $\rho$ is
the gas density.

To calculate Sobolev line optical depths, we assume that the atomic
level populations are set by local thermodynamic equilibrium (LTE).
This approximation should be reasonable in the optically thick regions
of ejecta, where the radiation field tends towards a blackbody
distribution.  In applying the a Sobolev formalism, we make two
further assumptions: first, that the intrinsic (Doppler) width of
lines is small compared to the velocity scale over which the ejecta
properties vary, and second, that the intrinsic profiles of strong
lines do not overlap with other lines.  While the first condition is
easily satisfied in rapidly expanding NSM ejecta, the second may not
be (see K13), and a non-Sobolev treatment may ultimately
be necessary for a fully rigorous treatment of the radiation
transport.

The expansion opacity takes a simplified form in atmospheres where
most lines are extremely optically thick ($\tau \gg 1$). As $\tau_i$
increases, the dependence of $\alpha_{\rm exp}$ on optical depth is
eliminated ($1 - e^{-\tau_i} \simeq 1$), and the expression for
expansion opacity simplifies to a sum of optically thick lines. The
dependence on density and other determinants of optical depth are
concomitantly reduced. Under these conditions, the number of distinct
optically thick lines in each bin becomes the most important predictor
of ejecta opacity.  An exhaustive tally of lines is therefore
essential to accurately modeling ejecta opacity.  Unfortunately, there
is relatively little line data available for the heavy elements ($Z >
28$) expected to be synthesized in NSM ejecta. We compiled the line
data provided in the VALD database \citep{VALD_2008}, which includes
fairly extensive data for a few heavy ions (e.g. CeII, CeIII), but
very little for most others species.

On theoretical grounds, we expect the lanthanides (atomic
numbers $Z = 58-72$) to contribute significantly to ejecta opacity,
due to the complicated structure of their valence f-shells. This
argument is illustrated with a simple combinatorics heuristic.  The
number of substates corresponding to a given electron orbital is
roughly
\begin{equation}
C = \frac{g!}{n! (g-n)!}~~{\rm with~}g = 2(2l +1 ),
\end{equation}
where $n$ is the number of valence electrons and $l$ is the angular
momentum quantum number of the valence shell.  The number of lines
should scale as $C^2$, and will be much greater for ions with
electrons in an open $f$ ($l=3$) shell.  Assuming lines from the two
species are equally likely to be optically thick, we expect
atmospheres containing lanthanides to have a much higher opacity than
atmospheres of pure iron. A similar argument could be applied to
elements of the actinide series, which may also be produced in the
merger ejecta. However, the mass fractions of most actinides are
predicted to be low, and can likely be ignored.

Initial calculations of opacities using line data from VALD suggested
that lanthanides have a much higher expansion opacity than iron (see
Figure \ref{Fig:1}). However, the limited line data available in
the VALD database makes it difficult to predict opacities over a range
of ionization states.
Instead, we use the theoretical lanthanide line data of
K13 derived from the atomic structure modeling code
\texttt{Autostructure} \citep{Badnell_2011}.  This supplies
approximate radiative data for neodymium ($Z = 60$) and a few other
elements over the entire wavelength range of interest ($ \sim 100
~\mathrm{\AA} - 30,000~ \mathrm{\AA}$).  We tested the reliability of
the \texttt{Autostructure} data by comparing predicted expansion
opacities of select species, including Fe and Ce, to those calculated
using existing line data in VALD.
Overall, the results from both data sets are found to be in good
agreement (Figure \ref{Fig:1}).

Until \texttt{Autostructure} models for all lanthanides are
calculated, we use an averaging scheme to predict ejecta opacities
based on a few representative species.  We recast the ejecta
composition computed by \citet{Roberts_2011} in terms of two groups of
elements: iron-like elements (i.e., the d-block of the periodic table)
and neodymium-like elements (i.e., the lanthanides).  The line data
for iron is taken from \citet{Kurucz_CD23}, while that of Nd is derived from
\texttt{Autostructure} calculations.  We take the mass fraction of Nd (Fe) to be
equal to the average mass fraction of the lanthanides (d-block
elements). The remainder of the composition is taken to be calcium,
which serves as a low-opacity filler with an appropriate ionization potential.

Since each ion species in the ejecta has a unique set of strong lines,
and since $\alpha_{\rm exp}$ increases with the number of strong
lines, mixtures with a greater diversity of
elements will have higher opacity.  Since both Nd and Fe both have
intermediate complexity for their respective blocks, their opacities
can presumably be taken as representative of other elements in the
same region of the period table.
We therefore assume that each lanthanide or d-block element in the
original compositions provides an opacity equal to that of Nd or Fe,
respectively, and arrive at a generalized expression for expansion
opacity
\begin{equation}
\alpha_{exp} = \frac{1}{ct_{\rm exp}} \sum\limits_{Z}N_{Z}\sum\limits_{i} \frac{\lambda_i}{\Delta \lambda_{c}}\left( 1 - e^{-\tau_{i}\left( \rho_Z\right)}\right),
\end{equation} 
where $Z$ runs over the representative elements (here Fe and Nd),
$N_Z$ is the number of elements in the block represented by $Z$, and
$\rho_Z = \chi_Z \rho$, with $\chi_Z$ the mean mass fraction of the
representative elements. Since lanthanide contributions dominate the
opacity, the boosting procedure effectively increases the opacity by a
factor of 14, the number of lanthanide species.
  
Figure~\ref{Fig:1} shows the expansion opacity calculated for typical
parameters of NSM ejecta ($\rho = 10^{-13} ~ {\rm g ~ cm^{-3}}, ~ T =
5 \times 10^3 {\rm ~K, ~ and }~ t_{\rm exp} = 1 ~\rm{day}$).  The
values vary with temperature and density, but in general our
calculations predict \rp\ opacities many orders of magnitude higher
than those calculated for iron group elements.

\begin{table*}
\caption{Peak magnitudes and bolometric light curve properties of radioactive transients}
\label{tab:sum}
\small
\centering
\begin{tabular}{ c c c | c c c c c c c}
\hline 
$M_{\rm ej} (M_\odot)$ & $\beta$ & composition & $L_{\rm p,bol}$\tablenotemark{a} & $t_{\rm p,bol}$\tablenotemark{b} & $M_B$ & $M_R$ & $M_I$ &$M_H$  \\
\hline \hline
$10^{-3}$ & $0.1$  & r-proc & $1.2 \times 10^{40}$ & 0.65  & -9.3 & -11.2 & -12.0 & -14.0  \\
$10^{-3}$ & $0.2$ & r-proc &$1.6 \times 10^{40}$ & 0.75 &  -9.1 & -11.3 & -12.6 & -14.1 \\
$10^{-3}$ & $0.3$ & r-proc & $3.5 \times 10^{40}$ & 0.15 &  -9.3 & -12.4 & -13.6 & -14.7 \\
$10^{-2}$ & $0.1$ & r-proc & $5.2 \times 10^{40}$ & 4.25 &  -11.2 & -12.7 & -13.3 & -15.8 \\
$10^{-2}$ & $0.2$ & r-proc & $8.5 \times 10^{40}$ & 1.85 &  -11.4 & -12.8 & -14.0 & -15.9 \\
$10^{-2}$ & $0.3$ & r-proc & $1.7 \times 10^{41}$ & 0.25 &  -11.8 & -14.1 & -15.2 & -16.5 \\
$10^{-1}$ & $0.1$ & r-proc & $2.4 \times 10^{41}$ & 8.65 &  -12.9 & -14.1 & -14.6 & -17.5 \\
$10^{-1}$ & $0.2$ & r-proc & $4.1 \times 10^{41}$ & 4.35 &  -13.4 & -14.3 & -15.4 & -17.5 \\
$10^{-1}$ & $0.3$ & r-proc & $7.2 \times 10^{41}$ & 0.25 &  -14.0 & -15.5 & -16.7 & -18.1 \\
$10^{-3}$ & $0.1$ & $^{56}$Ni  & $3.5 \times 10^{40}$ & 0.25 &  -13.0 & -12.6 & -12.6 & -11.8 \\
$10^{-2}$ & $0.1$ & $^{56}$Ni  & $3.7 \times 10^{41}$ & 0.75 &  -15.4 & -14.9 & -14.5&  -14.8\\
\hline \hline
\tablenotetext{1}{Peak bolometric luminosity, in ${\rm ergs~s^{-1}}$}
\tablenotetext{2}{Rise time to bolometric light curve peak, in days}
\end{tabular}
\end{table*}

\section{Light curves of \emph{r}-process transients}

The surprisingly high opacity of \rp\ material, discussed in
\S\ref{sec:rpop}, has important implications for the EM emission from
NSMs.  In this section, we present radiation transport calculations
using our refined opacity estimates to determine the bolometric and
broadband light curves of  \rp\ outflows.  Our
predictions diverge from those of earlier studies, which assumed that
the opacities were similar to those of iron.

\subsection{Ejecta Model}

We model the NSM ejecta as a spherically symmetric outflow undergoing
homologous expansion. In reality, the ejecta may have a highly
asymmetric, ``tidal tail" geometry \citep{Rosswog_2005}.  3-D
transport calculations suggest that this asphericity makes the
emission moderately anisotropic, but does not qualitatively
change the shape of the light curves \citep{Roberts_2011}.  We
describe the density of the ejecta using the broken power law
profile introduced by \citet{Chevalier_1989}, in which density
decreases as $r^{-\delta}$ in the inner layers of the atmosphere and
as $r^{-n}$ (with $n > \delta$) in the outer layers. The shift from
$n$ to $\delta$ occurs at the transition velocity
\begin{equation}
v_t = 7.1 \times 10^8 \zeta_v \left( E_{51}/M \right)^{1/2} {\rm ~ cm ~ s^{-1}}, 
\end{equation}
where $E_{51}$ is the explosion energy $E/10^{51}$ ergs, $M$ is the
ejecta mass $M_{\rm ej}$ in units of \Msun, and $\zeta_v$ is a numerical
constant. For $v < v_t$, the density is given by

\begin{equation}
\rho(r,t) = \zeta_{\rho} \frac{M_{\rm ej}}{v_t^3 t^3} \left( \frac{r}{v_t t} \right) ^{-\delta},
\end{equation}
and an analogous expression describes the outer layers. The constants
$\zeta_v$ and $\zeta_{\rho}$ satisfy the requirement that the density
profile integrates to the specified mass and energy.  We also
tried using an exponentially decreasing density profile, but found
that the light curves were mostly insensitive to the details of the density
structure.  The results presented here were generated using the broken
power law profile with $(n, \delta) = (1, 10)$.

The mass and velocities of the matter ejected during compact object mergers is
still uncertain. Simulations of NSMs \citep[e.g.][]{Rosswog_1999,
  Oechslin_2007, Goriely_2011, Hotokezaka_2013} and of neutron
star-black hole mergers \citep [e.g.][]{Janka_1999, Lee_2001,
  Rosswog_2005} have found that the amount of material ejected is
sensitive to binary type, mass ratio, nuclear equation of state, and the treatment
of gravity (Newtonian v. general relativistic).  However, taken in
aggregate, simulations suggest NSMs may eject $\sim 10^{-2} ~
M_{\odot}$ of material with velocities on the order of $\beta = v/c
\sim 0.1$.  We adopt these parameters as our fiducial model, but vary
the mass and velocity scales over the range $M = 10^{-3} -
10^{-1}M{\odot}$, and $\beta_{\rm char} = 0.1- 0.3$, where we define
the characteristic velocity by
\begin{equation}
E = \frac{1}{2}M_{\rm ej}\left( \beta_{\rm char} c \right) ^2.
\end{equation}
We assume the ejecta to be homogenous, and composed of either pure
\rp\ material or pure $^{56}$Ni.  For models with \rp\ elements, all
zones are assumed to have the same radioactive decay rate given by
 \cite{Roberts_2011}.  Of the decay energy, 10\% is taken
to be in fission fragments, and 90\% in beta decays.  Of the beta
decay energy, 75\% is assumed to be lost to neutrinos, with the
remaining 25\% split equally between leptons and gamma-rays.  Leptons
and fission fragments are assumed to be thermalized locally, while we
use a radiation transport scheme to follow the propagation and
absorption of gamma rays.  This approximate apportioning of the decay
energy is based on the physical considerations given in
\cite{Metzger_2010}.

\begin{figure}
%\vspace{-10pt}
\includegraphics{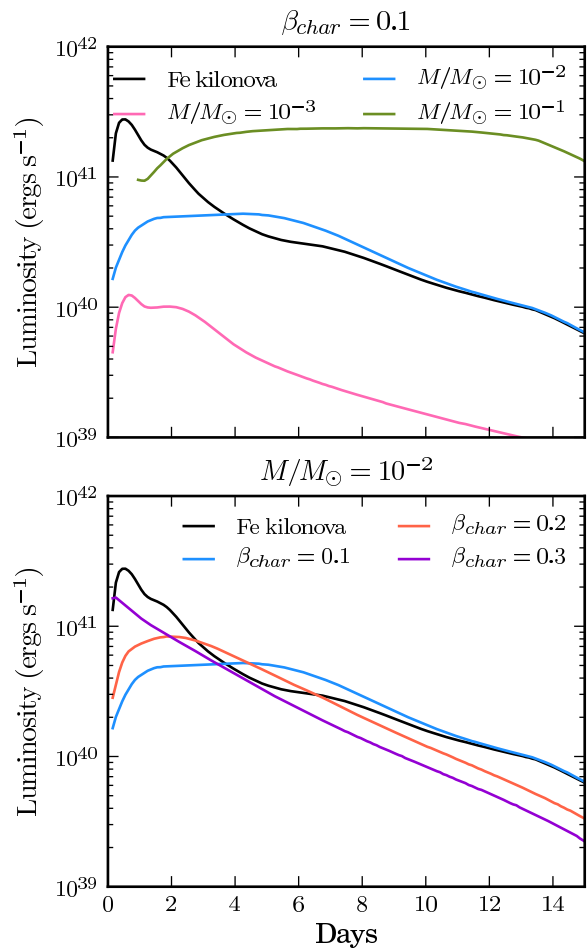}
\vspace{-20pt}
\caption{ Synthetic bolometric light curves of \rp\ transients with different ejecta masses (top panel) and velocities
(bottom panel). For comparison,
we also show the  light curve of a model with  fiducial ejecta parameters ($M = 10^{-2}~\Msun,~\beta_{\rm char} = 0.1$) but calculated assuming iron-like opacities (black lines).  The higher opacities of \rp\ ejecta lead to significantly broader light curves.  The models with higher ejecta velocities correspond to shorter rise times and steeper declines, while those with higher masses have greater luminosities and longer durations. 
 \label{Fig:2}}
 \end{figure}

\subsection{Light Curves}

\begin{figure}
%\vspace{-15pt}
\includegraphics{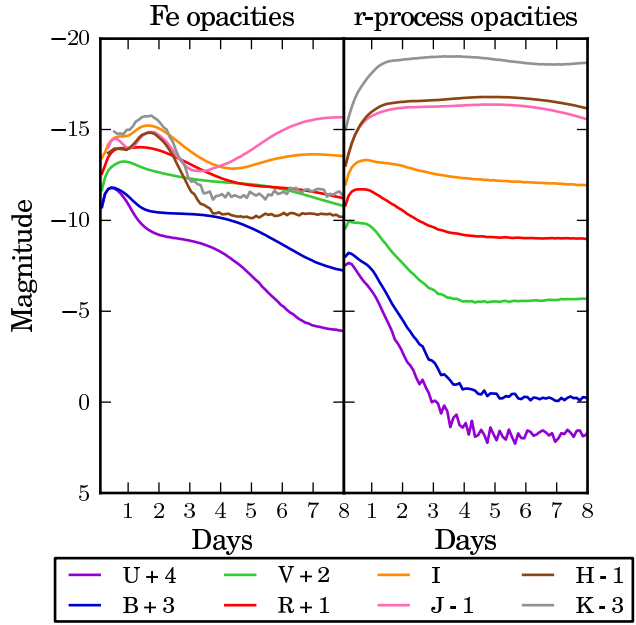}
\caption{ Synthetic broadband light curves calculated for the fiducial ejecta model calculated using
iron-like opacities (left) and \rp\ opacities (right).  The effect of \rp\ opacities is to suppress the optical emission and shift the radiation  toward redder bands, in particular the infrared J, H and K bands.}
\label{Fig:3}
\end{figure}

We  generate synthetic observables of our ejecta models using the
time-dependent multi-wavelength radiation transport code
\texttt{Sedona} \citep{Kasen_MC}.  Beginning at an initial time of
$0.1$ days after mass ejection, the code follows the temperature and
density evolution of the expanding ejecta, taking into account
radiative and radioactive heating as well as cooling by expansion.
The wavelength dependent \rp\ and iron opacities are calculated in
each zone using the \texttt{Autostructure} and \citet{Kurucz_CD23}
line data, respectively.  The ionization and excitation state of the
gas are assumed to be set by LTE, and lines are taken to be completely
absorbing.  \texttt{Sedona} synthesizes the emergent spectral time
series, from which we construct bolometric and broadband light curves.
Table \ref{tab:sum} summarizes the EM properties of the models we
investigated.

In Figure \ref{Fig:2}, we plot the bolometric light curves of
\rp\ transients for a subset of the parameter grid.  For comparison,
we also show a light curve calculated using iron-like opacities.
Introducing more realistic \rp\ opacities clearly changes the picture
significantly -- the light curves have slower rise times and broader,
dimmer peaks than those calculated with iron-like opacities. Notably,
the \rp\ transients do not in general have the $\sim 1 {\rm ~ day}$
duration thought to be a distinguishing feature of these events, and
instead may last a week or longer.

That higher opacities give broader, dimmer light curves is expected
theoretically. The duration of radioactively powered bolometric light
curves is set by the effective diffusion time through the ejected
material \citep{Arnett_1979},
\begin{equation}
t_{\rm d} \simeq \left( \frac{M_{\rm ej} \kappa}{v c} \right)^{\frac{1}{2}},
\label{eq:diff}
\end{equation}
where $v$ is a characteristic ejecta velocity and $\kappa$ an
appropriately wavelength-averaged opacity.  Because the \rp\ opacities
are $10-100$ times larger than those of iron, the diffusion time is
significantly lengthened.  The longer diffusion time also leads to a
dimmer luminosity at peak, as a greater fraction of the radioactive
energy is lost due to expansion before it can be radiated.  The models
roughly obey Arnett's law -- i.e., the emergent luminosity and
instantaneous radioactive energy input are approximately equal at peak
\citep{Arnett_1979, Arnett_1982}.

The \rp\ light curves vary with the ejecta properties in predictable
ways.  Higher mass ejections give a greater luminosity and longer
duration, due to their larger radioactive mass.  Ejecta models with
higher kinetic energies have shorter rise times, reach greater peak
luminosities, and decline more rapidly than their lower energy
analogues.  Over the reasonable range of ejecta parameters considered
here, the light curves exhibit significant diversity -- the peak
luminosities vary by more than an order of magnitude, and the
durations range for $\lesssim 1$~day to as long as two weeks.

\begin{figure}
\vspace{-10pt}
\includegraphics{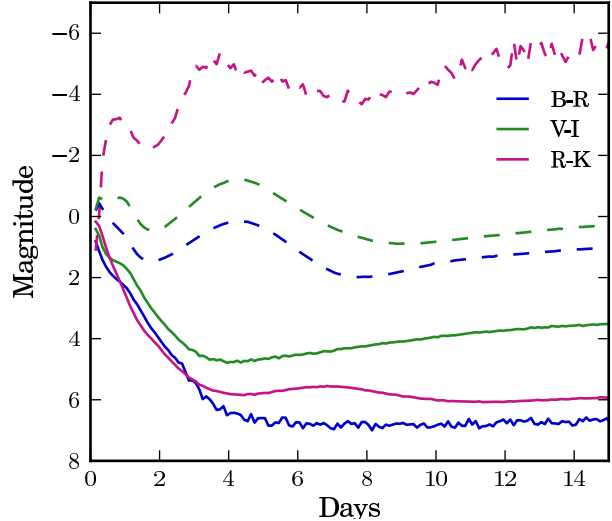}
\caption{ Select colors for the fiducial model calculated using \rp\ (solid lines) and iron-like (dashed lines) opacities. The colors of \rp\ transients initially redden rapidly, then reach a phase of constant
color characterized by a blackbody at $T \approx 2500$~K.}
\label{Fig:4}
\end{figure}

The broadband magnitudes of the models also
differ significantly from previous expectations. Figure \ref{Fig:3}
shows that, compared to a model that uses iron-like opacities,
\rp\ transients output much more energy in red and infrared bands,
with a strong suppression of the optical emission. We find bright, broad
peaks in the J, H, and K bands, while the U, B, and V bands are
heavily line blanketed and decline sharply at early times.  Figure
\ref{Fig:4} shows that the colors of the fiducial \rp\ model redden
rapidly over the first day or two, and afterwards become remarkably
constant, with the SED peaking in the infrared at around $\sim
1~\mu$m. Other than the very red color, the spectra at these phases
resemble those of other high-velocity SNe, with a pseudo blackbody
continuum and broad ($\sim 200$~\AA) spectral features (see K13).

\begin{figure}
\includegraphics{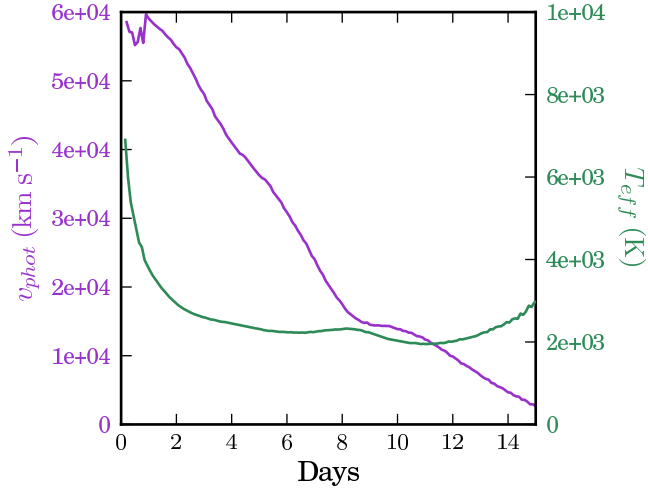}
\vspace{-20pt}
\caption{Evolution of the photospheric temperature and velocity in the fiducial model, using \rp\ opacities.  The photospheric temperature initially declines rapidly, but at times $\gtrsim 2$ days settles to a fixed value  near the
recombination temperature of the lanthanides.   }
\label{Fig:photofig}
\end{figure}

The behavior of the broadband light curves can be understood by
examining the photospheric properties of \rp\ transients, since the
observed SED roughly corresponds to a blackbody at the photospheric
temperature and radius.  In Figure \ref{Fig:photofig}, we plot the
velocity and temperature evolution of the photosphere, the surface
defined by
\begin{equation}
\tau(r_{\rm phot}) = -\int_{\infty}^{r_{\rm phot}} \overline{\kappa}_P(r) \rho(r) \mathrm{d}r = 1,
\end{equation}
where  $\overline{\kappa}_P(r)$ is the Planck mean opacity, computed from our 
\rp\ line data.
%\begin{equation}
%\overline{\kappa}_P(T) = \frac{\int_0^{\infty} \kappa_{\nu} ~ B_{\nu} (T ) \mathrm{d}\nu}{\int_0^{\infty} B_{\nu}(T) \mathrm{d}\nu}
%\end{equation}

In the initial phases, the photospheric velocity and temperature
decline steadily, reflecting the decrease in density and cooling
of the ejecta due to expansion.  At around $\sim {\rm 2 ~
  days}$, however, the photospheric temperature stabilizes at $T
\simeq \rm{2500 ~ K}$. Since this is close to the first ionization
temperature of the lanthanides, this plateau probably reflects the
sharp drop in opacity that occurs when these elements recombine to
neutral (see K13).  Recombination occurs in the cooler
outer layers first, and a sharp ionization front forms in the ejecta.
Photons pass easily through the cooler, neutral outer layers, but are trapped
in the ionized inner regions -- the
photosphere thus forms at the ionization front.  During this phase, the emergent colors are
roughly constant in time and resemble those of a blackbody at the
lanthanide recombination temperature.  Over time, the recombination
front recedes inward, reaching the center at around 14~days.  At this
point, the ejecta is nearly entirely neutral and transparent.

\subsection{Uncertainties in the Opacities}

\begin{figure}
\vspace{-10pt}
\includegraphics{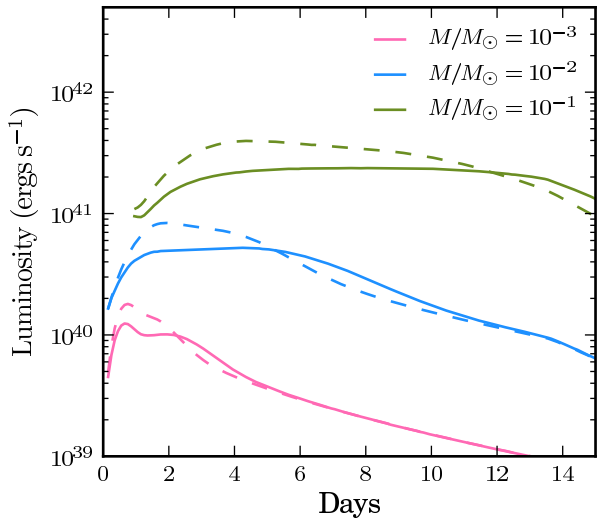}
\caption{Bolometric light curves for $\beta_{\rm char} = 0.1$ and a range of ejecta masses, calculated using line data with from two different \texttt{Autostructure} models of Nd, each with a somewhat different energy level structure. Solid curves correspond to our preferred \texttt{Autostructure} model ({\it opt3}), while dashed curves show results from an alternative model ({\it opt2}).  The moderate differences can be taken as a rough estimate  of the error resulting from  uncertainties in the Nd structure models.
\label{Fig:5}}
\end{figure}

\begin{figure}
\vspace{-10pt}
\includegraphics{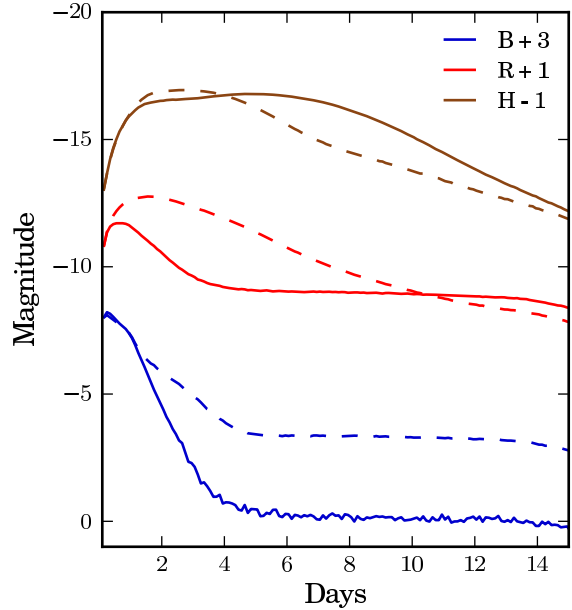}
\caption{A comparison of broadband curves of the fiducial model,
calculated using line data with from two different \texttt{Autostructure} models of Nd, each with a somewhat different energy level structure.  Calculations using the {\it opt2} line data (dashed lines) predicts higher magnitudes for the optical B-, and R-bands than those with the {\it opt3} line data (solid lines)}
\label{Fig:6}
\end{figure}

Though our \rp\ opacities represent an improvement over previously available  data, the \texttt{Autostructure} models of Nd are subject to  uncertainties. In particular,  the structure models rely on an \textit{ab-initio} optimization, such that the predicted atomic level energies and line wavelengths generally differ from the experimental values.  To explore the effects of these uncertainties, we calculated bolometric and broadband light curves using two different \texttt{Autostructure} models of Nd ({\it opt2} and {\it opt3} from K13) each with a somewhat different energy level structure.  The {\it opt3} model--which we have used in the calculations presented thus far--reproduces the low lying energy levels of NdII quite well, while the {\it opt2} model has generally higher excitation energies which are harder to populate under our assumption of LTE. The result is fewer strong lines and a lower overall opacity for the {\it opt2} case.

In Figure \ref{Fig:5}, we plot bolometric light curves for  ejecta models with $\beta_{\rm char} = 0.1$ and a variety of masses. 
The light curves calculated using the  {\it opt2} line data have somewhat sharper, more luminous peaks and swifter declines, consistent with the expected lower opacities. 
The differences, however, are fairly modest, and the bolometric luminosity never differs by more than a factor of $\sim 3$.     The effects are more noticeable in the broadband light curves (Figure \ref{Fig:6}).
In particular, the R-band light curves are $\sim 1 -2$ magnitudes brighter for the {\it opt2} data, 
which could have important implications for detectability.  Based on comparison to experiment, we expect the {\it opt3} data to be more reliable, however further work refining the opacities is clearly warranted.

In addition to errors inherent in the individual structure models, a perhaps larger uncertainty in our \rp\ opacities arrises from the fact that we represent all lanthanides  with the radiative data of Nd.  In fact, the lanthanides
with a nearly half open f-shell (in particular gadolinium) will be significantly more complex, with perhaps $\sim 10$ times as many lines of Nd.  We therefore suspect that our current calculations  underestimate the true opacity.   The light curves we present here  suggest that  increasing the opacity further will
lead to an even greater suppression of optical emission.

\subsection{A \Nifs-powered transient}

\begin{figure}
\includegraphics[width=3.2in]{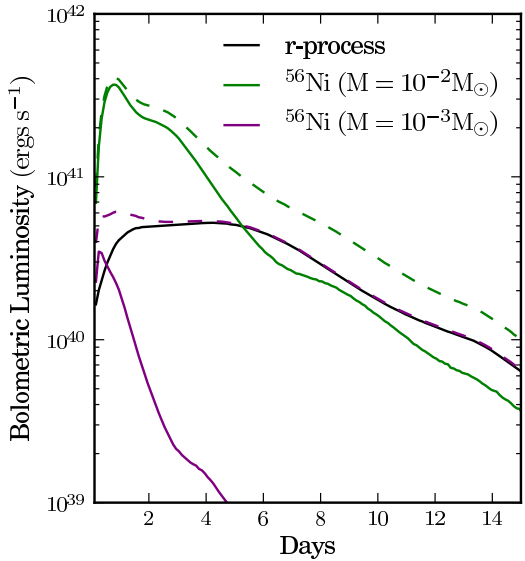}
\caption{Bolometric light curves for models that include two components: \rp\ material (from tidal tails) and \Nifs\ (from a disk wind).  We plot the light curve of the fiducial \rp\ ejecta model ($M_{\rm rp} = 10^{-2}~\Msun$, black line) along with
two models of pure \Nifs\ with different masses ($M_{\rm ni} = 10^{-2}~\Msun$ and $10^{-3}~\Msun$, green and purple solid lines).  The dashed lines
give the combined two-component light curves.}
\label{Fig:Ni}
\end{figure}

\begin{figure}
\includegraphics[width=3.2in]{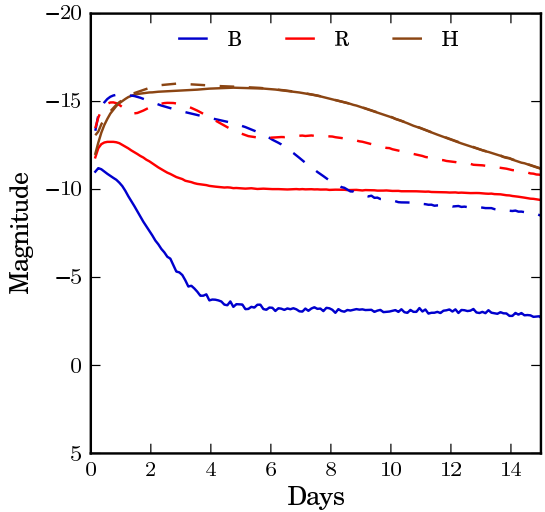}
\caption{ A comparison of select broadband light curves for a pure \rp\ transient (solid lines) and an \rp\ transient combined with a \textsuperscript{56}Ni-powered outflow (dashed lines). The bluer SED from the \textsuperscript{56}Ni shifts the magnitudes of the bluer bands of the combined SED upward relative to a pure \rp\ model. This plot is for $M_{\rm ni} = M_{\rm rp} = 10^{-2}M_{\odot}.$ }
\label{ni-cols}
\end{figure}

Given that the radioactive light curves of NSMs depend strongly on the composition
of the ejecta,  it is worth considering whether any elements lighter than the lanthanides may be 
produced in these events.
While the material dynamically ejected in the merger itself (the tidal tails) is thought to undergo
robust \rp\ nucleosynthesis,  it is 
plausible that a comparable amount of mass  may subsequently be blown off in winds from 
an accretion disk surrounding the merged remnant.
Though the physical properties of the disk winds remain uncertain,  neutrino irradiation 
may drive the electron fraction to $Y_e \gtrsim 0.4$, in which case the nucleosynthesis
may not extend past $Z \sim 50$  \citep{Surman_2006, Surman_2008, Metzger_2008, Darbha_2011}.  If $Y_e$ is very close to 0.5, the composition will
be primarily  \textsuperscript{56}Ni.
In this case, the EM signature of a merger may be a superposition of a \textsuperscript{56}Ni- and a \rp-powered transient.

To address this possibility, we consider a simplified scenario where $10^{-3}-10^{-2} M_{\odot}$ of pure \textsuperscript{56}Ni is blown off in a wind immediately post-merger. 
Consistent with our use of spherical symmetry thus far, we model this wind as a spherical outflow, with $\beta_{char} = 0.1$ and the same broken power law density profile with $(n, \delta) = (1,10)$.
We consider the tidal tails and disk wind to be two separate, non-interacting components, which is
perhaps not unreasonable given that the winds are likely collimated in the polar regions, while 
the tidal tails are largely confined to the orbital plane.  Ignoring viewing angle effects, we take the two component light curve to simply be the superposition of the individual \textsuperscript{56}Ni-powered and the \rp\ powered light curves.  

Figure \ref{Fig:Ni} shows the two component light curves, for two different ratios of the \Nifs\ wind mass ($M_{\rm ni}$) to the \rp\ tidal tail mass ($M_{\rm rp}$). For $M_{\rm ni} \ll M_{\rm rp}$, the primary effect of the \Nifs\ wind is to raise the early-time luminosity, creating a very short peak at $t  \sim {\rm 1 ~ day}$, which blends into the long, flat, \rp\ light curve. The cumulative light curve thus appears to have a faster rise time and longer plateau. If $M_{\rm ni} \approx M_{\rm rp}$, the \Nifs\ emission dominates the \rp\ emission for the first $\sim 5$ days post merger, with the two components contributing roughly equally thereafter. The net effect is a gradually
declining light curve, with the long \rp\ plateau obscured by the \Nifs-powered light curve.

\begin{figure}
\includegraphics[width=3.2in]{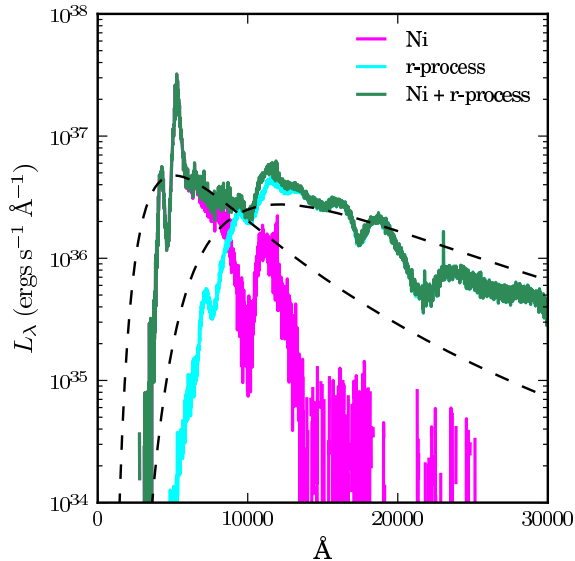}
\caption{ A combined \Nifs\ and \emph{r}-process spectrum at $t = 7$ days, taking  $M_{\rm ni} = M_{\rm rp} = 10^{-2}M_{\odot}.$ The peak at blue wavelengths is due to the \Nifs\, while the \rp\ material supplies the red and infrared emission. The best fit blackbody curves to the individual spectra are overplotted in dashed black lines ($T_{\rm ni} \simeq 5700$ K, $T_{\rm rp} \simeq 2400$ K). The combined spectrum roughly resembles the superposition of two blackbodies at different temperatures.}
\label{fig:nirp_spec}
\end{figure}

The addition of a \Nifs\ component also affects the SED of the transient, as shown in Figure~\ref{ni-cols}
for the case $M_{\rm ni} = M_{\rm rp} = 10^{-2}M_{\odot}.$     Given the much lower iron group opacities, the  SED of the \Nifs\ ejecta is much bluer than that of the \rp\ ejecta.  The emission in the optical bands (U,B,V,R)  is relatively bright and set by \Nifs\ mass, while  the \rp\ material establishes the behavior in the infrared bands. 
Such an unusual SED may serve as an EM fingerprint that could improve the prospects for positively identifying a NSM.  In particular, as shown in Figure~\ref{fig:nirp_spec}, the spectrum of a two component outflow is, to first approximation,  the superposition of two blackbodies -- a sharply peaked bluer blackbody, corresponding to the \Nifs\ ejecta, and a lower, redder one, corresponding to the \rp\ material.

The aggregate light curve model we present here glosses over some of the more complex physical processes. Our model assumes spatially distinct regions of pure \textsuperscript{56}Ni and pure \rp\ material. In reality,
the nucleosynthetic yields are highly sensitive to the conditions in the wind, and it is possible that disk outflows contain some elements heavier than \Nifs.
Contamination of the outflows with even a small mass fraction of lanthanides ($\sim 10^{-3}$) can
significantly increase the opacities and the optical line blanketing.
Even if our simplified compositions  turn out to be reasonable, our model does not account for the geometry of the ejecta and any possible mixing of the wind and tidal tail components.  Given the
presumably high level of asymmetry, the net (tails + wind) EM output may depend heavily on orientation, making our simple superposition procedure valid only along certain lines of sight.

\section{Conclusion}

We have shown that the  radioactive powered light curves associated with NSMs are 
greatly modified when more realistic values for the opacities of \rp\ material are taken into account.
The \rp\ opacities are much higher than those of iron, due to both the complexity of heavy elements (in particular the lanthanides) and the diversity of atomic species present.  Refining our understanding of the atomic structure of these elements is an important step toward a more rigorous model of transients from merging compact objects.

In accordance with theoretical expectations, the extremely high \rp\ opacities result in bolometric light curves that are broader and dimmer than those calculated assuming iron-like opacities. 
Our calculations indicate that the light curves are likely to last at least a few days, and may endure as long as a week or two  in certain cases.  The broadband magnitudes are also significantly impacted -- we find heavy line blanketing in the optical and UV bands, with most of the radiation emitted in the near infrared.  The colors at later
times are fairly constant, and regulated to be similar to a blackbody at $T \approx 2500$~K, the recombination 
temperature of the lanthanides.

These findings have important, if mixed, consequences for the detectability of  EM counterparts to NSMs. On the one hand, we predict dimmer bolometric luminosities and  SEDs  largely  shifted into the infrared, both of which pose serious challenges to observational surveys at optical wavelengths.   On the other hand, the light curves are of longer duration, and may not require quite
as a high cadence of observations.   Perhaps more importantly,  the uniquely high opacity of \rp\ ejecta
 provides  signatures that may allow us to distinguish NSMs from other sorts
of dim transients.  In particular, the SED of \rp\  ejecta peaks in the infrared, with a color temperature set by
lanthanide recombination.   If the merger
ejects two separate mass components -- \rp\ tidal tails and a \Nifs\ wind -- the dual spectrum
may be quite distinctive, with discernible infrared and optical components.

The SEDs we predict can be used to roughly estimate the detectability, given the varying depths and wavelength coverage of different observing facilities \citep[e.g.,][]{Nissanke_2012}. For example, Pan-STARRS (see http://pan-starrs.ifa.hawaii.edu) and PTF \citep{Law_2009_PTF} achieve an R-band depth of $M_R \sim$ 21 magnitudes, while LSST reaches a depth of $M_R \sim$ 24 \citep{LSST_09}. 
We find that an \rp\ transient with fiducial model parameters will peak at $M_R = -13$, which under ideal observing conditions, would be observable to Pan-STARRS or PTF out to a distance of $\sim$ 60 Mpc.  This
is an interesting, but rather small fraction of the  volume probed by advanced LIGO/VIRGO.
The case with LSST is more promising, with sensitivity in the R-band out to $\sim 250$~Mpc.
Discovery of \rp\ ejecta in the U or B bands with any facility would appear to be quite difficult, given the heavy line blanketing at these wavelengths.

Given that our models predict that most of the emission is at longer wavelengths, improving detection capabilities in the near infrared may greatly aid in future searches for EM counterparts.  
Ground based facilities with sensitivity in the I or Y bands ($0.8 - 1.1 \mu$m) may benefit from these 
capabilities, as the \rp\ transients are generally $\sim 1$ magnitude brighter in these bands than in R-band.  
The  construction of space based facilities such as WFIRST \citep{Green_2012_WFIRST} and Euclid \citep{Amendola_2012_Euclid} would be of particular interest. WFIRST is proposed to have an H-band depth of $\sim$ 25 magnitudes, with Euclid achieving a similar sensitivity.  As our fiducial model is much brighter in the infrared ($M_{H} \simeq$ -15) than in the optical bands, such facilities could potentially make a detection out to a distance of $\sim$ 1000 Mpc, encompassing the entire  LIGO/VIRGO volume.

Discovering the EM counterparts to NSMs would be made significantly easier if, in addition to \rp\ elements, these events also separately eject some significant amount of \Nifs\ or lower mass ($Z < 58$) radioactive isotopes.  Our models predict that such ``lanthanide-free'' light curves  are reasonably bright in the optical bands ($M_B \approx M_R \approx -15$)  and  would be within range for many upcoming 
optical transient surveys.   It is plausible that winds from a post-merger
accretion disk may produce such lighter element outflows, although more detailed simulations are needed to 
constrain the mass and composition of the material ejected.  Clearly any  
detection of a short-lived optical transient should, if possible, be immediately followed up at infrared wavelengths to look for a coincident \rp\ transient from the tidal tails.  Discovery 
of such a two component light curve and spectrum would be a very strong signature of a NSM. 
It would also provide  insight into the merger and post-merger physics by separately constraining
the mass ejected by different mechanisms.

The work we have presented here is an important step towards improving our predictions
of the radioactive transients from NSMs. However, much  remains to be done.  The opacities
we have used, while more realistic than previous estimates, are still subject to important uncertainties.
In particular, we need comprehensive structure calculations to derive radiative data for all lanthanides. 
In addition, more detailed simulations of the dynamics and nucleosynthesis of the mass ejection are needed to better predict the mass, composition, and geometry
of the ejecta.  Of special interest are the  properties of the disk wind and any mixing of these outflows with the tidal tail material.
Finally, 3-dimensional radiative transfer calculations will be needed to predict
the light curves of multi-component mass ejections and to determine their dependence on viewing angle.   
Such theoretical work should improve our understanding of the EM counterparts to gravitational
wave sources, and the heavy elements they  produce.

\begin{acknowledgements}
This research has been supported by a Department of Energy Office of Nuclear
Physics Early Career Award, and by the Director, Office of Energy
Research, Office of High Energy and Nuclear Physics, Divisions of
Nuclear Physics, of the U.S. Department of Energy under Contract No.
DE-AC02-05CH11231.  This work is supported 
in part by an NSF Division of Astronomical Sciences collaborative 
research grant AST-1206097.
We are grateful for computing time made available
the National Energy Research Scientific Computing Center, which is supported by the Office of Science of the U.S. Department of Energy under Contract No. DE-AC02-05CH11231. 
\end{acknowledgements}

\end{document}